\documentstyle[12pt]{article}

\def\be{\begin{equation}}
\def\ee{\end{equation}}
\def\ba{\begin{eqnarray}}
\def\ea{\end{eqnarray}}

\def\asd{anti-self-dual}
\def\hk{hyperk\"ahler}

\def\pdes{partial differential equations}
\def\P{Pleba{$\tilde {\rm n}$}ski}

\def\sd{self-dual}
\begin{document}
\begin{titlepage}
\rightline{DAMTP-R92/47}
\rightline{December 1992}
\vskip 1.5cm
\centerline{\LARGE On Self-Dual Gravity}
\vskip 1cm
\centerline{\large James D.E. Grant
\normalsize\footnote{email address: jdeg1@phx.cam.ac.uk.}}
\vskip 0.5cm
\centerline{Department of Applied Mathematics and Theoretical
Physics,}
\vskip2pt
\centerline{University of Cambridge, Silver Street,
Cambridge CB3 9EW, United Kingdom.}
\vskip 1cm
\begin{abstract}
We study the Ashtekar-Jacobson-Smolin equations that
characterise four dimensional complex metrics with {\sd} Riemann
tensor. We find that we can characterise any {\sd} metric by a
function that satisfies a non-linear evolution equation, to
which the general solution can be found iteratively. This formal
solution depends on two arbitrary functions of three
coordinates. We construct explicitly some families of solutions
that depend on two free functions of two coordinates, included
in which are the multi-centre metrics of Gibbons and Hawking.
\end{abstract}
\end{titlepage}

\section{Introduction}

In four dimensions the Hodge duality operation takes two forms
to two forms. Given a four dimensional metric, the most
important two form associated with it is the curvature two form
$R^a_{\ b}$. It is therefore natural to be interested in four
dimensional metrics whose curvature form obeys the self-duality
relation
\be
R^a_{\ b}\,=\,^{*}R^a_{\ b},
\ee
where $^{*}$ is the Hodge duality operator. We will refer to
such metrics as ``\sd''. Such metrics automatically have
vanishing Ricci tensor, and so satisfy the vacuum Einstein
equations with vanishing cosmological constant. Unfortunately,
the only real Lorentzian {\sd} metric is flat Minkowski space,
so we choose instead to work with metrics with four {\it
complex} dimensions.

Physically these metrics may be of interest in attempts to
quantise gravity, since they correspond to saddle points of the
Einstein-Hilbert action, therefore giving large contributions to
a path integral over euclidean metrics \cite{SWH1}.
Alternatively, it may be possible to interpret them as ``one
particle states'' in a quantised gravity theory \cite{NLG}.

{}From a purely mathematical point of view these metrics are
interesting since they are ``\hk''. Hyperk{\"a}hler manifolds are
$4n$ dimensional manifolds (for $n$ a positive integer) that
admit a non-singular euclidean metric, ${\bf g}$, with respect to
which there exist three automorphisms, ${\bf J}^i$, of the
tangent bundle which obey the quaternion algebra \cite{CAL}. In
other words
\be
\nabla\,{\bf J}^i\,=\,0,\,{\bf J}^i\,{\bf
J}^j\,=\,-\,{\delta}_{ij}\,+\,{\epsilon}_{ijk}\,{\bf J}^k,
\ee
where $\nabla$ is the covariant derivative with respect to the
metric ${\bf g}$. In four dimensions, it turns out that for a
metric, ${\bf g}$, to be {\hk} it must have either \sd, or
\asd, curvature tensor \cite{ATHISI}.

The problem of constructing metrics with {\sd} curvature tensor
has been tackled in several ways. The most direct approach is to
formulate the problem in terms of {\pdes} \cite{PLEB,NEWM}. A
more constructive approach is Penrose's `Non-Linear Graviton'
technique \cite{NLG}. Here, the task of solving {\pdes} is
replaced by that of constructing deformed twistor spaces, and
holomorphic lines on them. In practise this turns out to be just
as difficult as solving \pdes, but in principle one can
construct the general {\sd} metric in this way.

Here we concentrate on \pdes. We find a formulation which is
similar to \P's First Heavenly equation \cite{PLEB}, but which
can be viewed as simply an evolution equation. This means that
the free functions in our solution are just a field and its time
derivative on some initial hypersurface i.e. two free functions
of three coordinates. We construct, in a somewhat formal manner,
the general solution to this equation. We also construct
explicitly some infinite dimensional families of solutions to
these equations. In the appendices, we show how this formulation
is equivalent to \P's.

\section{Construction of Self-Duality Condition}

In \cite{AJS} the equations for complex {\sd} metrics were
reformulated in terms of the new Hamiltonian variables for
General Relativity introduced in \cite{ASH}. By fixing the four
manifold to be of the form ${\cal M} = {\Sigma} {\times} R$ and
using the coordinate $T$ to foliate the manifold, they reduced
the problem of finding {\sd} metrics to that of finding a triad
of complex vectors $\{{\bf V}_i:i=1,2,3\}$ that satisfy the
equations
\be
Div\, {\bf V}_i = 0,
\label{vone}
\ee
\be
{{\partial} \over {\partial T}}\,{\bf V}_i =
{1\over 2}\epsilon_{ijk}\,[{\bf V}_j,{\bf V}_k].
\label{vtwo}
\ee
Defining the densitised inverse three metric
\be
{\hat q}^{ab}\,=\,V_i^a\,V_j^b\,{\delta}_{ij},
\ee
we recover the undensitised inverse three metric $q^{ab}$ by
the relation $q^{ab} = {\hat q}\,{\hat q}^{ab}$, where ${\hat q}
= \det {{\hat q}_{ab}} = (\det {{\hat q}^{ab}})^{-1}$. If we now
define the lapse function $N$ by $N = ({\det q_{ab}})^{1/2}$
then we find that the metric defined by the line element
\be
ds^2 = N^2\,dT^2 + q_{ab}\,dx^a\,dx^b
\label{sedume}
\ee
is \sd.

Later, it was found that this triad of vectors could be related
to the complex structures ${\bf J}^i$ that {\hk} metrics admit
\cite{ROB}. Given a {\sd} metric, we choose local coordinates
$(T,x^a)$ to put the line element in the form of equation
(\ref{sedume}). If we define the triad of vectors ${\bf
V}_i\,=\,-\,{\bf J}^i(*,{\partial}_T)$, then these vectors will
satisfy (\ref{vone}) and (\ref{vtwo}).

Here we will concentrate on the problem of finding {\it local}
solutions to equations (\ref{vone}) and (\ref{vtwo}). We thus
introduce a local coordinate chart $(X,Y,Z)$ on the three
surface, $\Sigma$, with its natural flat metric and connection.
Thus (\ref{vone}) becomes just
\be
{\partial \over {\partial x^a}} V^a_i = 0.
\ee
The crucial step is to realise that we can write equation
(\ref{vtwo}) as
\be
[{\partial \over {\partial T}},{\bf V}_i]
= {1\over 2}\,\epsilon_{ijk}\,[{\bf V}_j,{\bf V}_k].
\label{clever}
\ee
If we consider only euclidean metrics, then we take the ${\bf
V}_i$ to be real. In this case we define two complex vectors
${\bf A}, {\bf B}$ by
\be
{\bf A} = {\partial \over {\partial T}}
+ i\, {\bf V}_1,
\ee
\be
{\bf B} = {\bf V}_2 - i\, {\bf V}_3.
\label{cvvec}
\ee
which, by virtue of (\ref{clever}), obey the Lie bracket algebra
\be
[{\bf A}, {\bf B}] = 0,\
[{\bf {\bar A}}, {\bf {\bar B}}] = 0,\
[{\bf A}, {\bf {\bar A}}] + [{\bf B}, {\bf {\bar B}}] = 0,
\ee
where $\ {\bf {\bar {}}}\ $ denotes complex congugate. We can
generalise these equations by considering four complex vectors
${\bf U},{\bf V},{\bf W}$ and ${\bf X}$ that satisfy the
relations
\be
[{\bf U}, {\bf V}]\,=\,0,
\label{liea}
\ee
\be
[{\bf W}, {\bf X}]\,=\,0,
\label{lieb}
\ee
\be
[{\bf U}, {\bf W}]\,+\,[{\bf V}, {\bf X}]\,=\,0.
\label{liec}
\ee
Here we are thinking of ${\bf W}$ and ${\bf X}$ as ``generalised
complex conjugates'' of ${\bf U}$ and ${\bf V}$ respectively. By
Frobenius' theorem, we can use (\ref{liea}) to define a set of
coordinates $(t, x)$ on the 2 (complex) dimensional surface
defined by vectors ${\bf U}$ and ${\bf V}$, and take ${\bf U}$
and ${\bf V}$ to be
\be
{\bf U} = {\partial \over {\partial t}},\,
{\bf V} = {\partial \over {\partial x}}.
\label{coords}
\ee
We can now foliate our whole space using the coordinates
$(t,x,y,z)$. The equation (\ref{liec}) then becomes %
${\partial}_t {\bf W} + {\partial}_x {\bf X} = 0$.
This means there exists a vector field ${\bf Y}$ such that %
${\bf W} = {\partial}_x{\bf Y},
{\bf X} = -{\partial}_t {\bf Y}.$ %
Thus we are only left with the problem of solving for vectors
${\bf Y}$ that satisfy %
$[{\partial}_t{\bf Y},{\partial}_x{\bf Y}] = 0$%
\footnote{It was only after this work was completed that
I learned of \cite{CMPLUSN} where the ideas developed so
far were found independently. From here onwards, however, our
treatments are different.}. %
We expand ${\bf W}$ and ${\bf X}$ as
\be
{\bf W} =
{\partial}_t + f_x {\partial}_y + g_x {\partial}_z,
\label{wexp}
\ee
\be
{\bf X} = - f_t {\partial}_y - g_{t} {\partial}_z.
\label{xexp}
\ee
(The reason for the ${\partial}_t$ term in ${\bf W}$ is, as
alluded to above, we are thinking of ${\bf W}$ as a sort of
complex conjugate of ${\bf U}\,=\,{\partial}_t$. Although this
argument only seems sensible for $t$ a real coordinate, we are
still perfectly at liberty to expand ${\bf W}$ in this way if $t$
is complex.) If, by analogy with (\ref{vone}), we impose
${\partial \over {\partial x^a}} W^a = {\partial \over {\partial
x^a}} X^a = 0$, then we find that there exists a function
$h(t,x,y,z)$ such that $f=h_z, g=-h_y$. Imposing (\ref{lieb}), we
find that there exists a function $\alpha (t,x)$ such that
\be
h_{tt} +
h_{xz} h_{ty} - h_{xy} h_{tz} = \alpha (t,x).
\ee
We can absorb the arbitrary function $\alpha$ into the function
$h$, and conclude that we can form a {\sd} metric for any
function $h$ that satisfies
\be
h_{tt} + h_{xz} h_{ty} - h_{xy} h_{tz} = 0.
\label{biggy}
\ee
This is just an evolution equation. Thus we can arbitrarily
specify data $h$ and $h_t$ on a $t = constant$ hypersurface and
propagate it throughout the space according to (\ref{biggy}) to
get a solution. For example, if we expand $h$ around the $t=0$
hypersurface, and insist that $h$ is regular on this surface,
then $h$ is of the form%
\be
h = a_0\,(x,y,z)\,+\,
a_1\,(x,y,z)\,t\,+\,a_2\,(x,y,z)\,{t^2 \over {2!}}\,
+\,a_3\,(x,y,z)\,{t^3 \over {3!}}\,+\dots
\label{aitch}
\ee
Substituting this into (\ref{biggy}) shows that $a_0$ and $a_1$
are arbitrary functions of $x,\,y$ and $z$. $a_2,\,a_3 \dots$ are
then completely determined for chosen $a_0$ and $a_1$ by
\be
a_2\,=\,a_{0xy}\,a_{1z}\,-\,a_{0xz}\,a_{1y},
\ee
\be
a_3\,=\,a_{0xy}\,a_{2z}\,-\,a_{0xz}\,a_{2y}\,+
\,a_{1xy}\,a_{1z}\,-\,a_{1xz}\,a_{1y},
\ee
and so on. Thus, in principle, we have a solution that depends on
two arbitrary functions of three coordinates. It is interesting
to compare our equation (\ref{biggy}) with {\P}'s First Heavenly
equation
\be
{\Omega}_{p {\tilde q}}\,{\Omega}_{{\tilde p}q}\,-\,
{\Omega}_{p {\tilde p}}\,{\Omega}_{q{\tilde q}}\,=\,1.
\label{heaven}
\ee
Here it is not so obvious what our free functions are, and an
expansion along the lines of (\ref{aitch}) doesn't work. (It is
shown in the appendix how to get equation (\ref{heaven}) from
our equation, showing that the two approachs are equivalent.
Thus for {\it any} {\sd} metric there will exist a corresponding
function $h$ that satisfies (\ref{biggy}).)

{}From the work of \cite{MASNEW} we know that the vectors ${\bf
U},\,{\bf V},\,{\bf W},\,{\bf X}$ are proportional to a null
tetrad that determines a {\sd} metric. Indeed the tetrad is
given by ${\sigma}_a = f^{-1} {\bf V}_a$, where ${\bf V}_a =
({\bf U},\,{\bf V},\,{\bf W},\,{\bf X})$ for $ a =
0,\,1,\,2,\,3$ and $f^2 = {\epsilon}\, ({\bf U},\,{\bf V},\,{\bf
W},\,{\bf X})$, for $\epsilon$ the four dimensional volume form
$dt \wedge dx \wedge dy \wedge dz$. In our case, $f^2 = -h_{tt}$
and our line element is
\be
ds^2\, =\, dt\,(h_{ty}\, dy + h_{tz}\, dz)\, +\,
dx\,( h_{xy}\, dy\,+\, h_{xz}\, dz)\, +\, {1 \over {h_{tt}}}\,
(h_{ty}\, dy\, +\, h_{tz}\, dz)^2.
\label{metric}
\ee

\section{The Formal Solution.}

We now construct, at least formally, the general solution to
(\ref{biggy}). Instead of working with this equation directly,
it is helpful to define two functions $A = h_t,\,B = h_x$, and
rewrite (\ref{biggy}) in the equivalent form
\be
A_t + A_y B_z - A_z B_y = 0,
\label{AH}
\ee
\be
A_x = B_t.
\label{BEE}
\ee
If we just viewing $B$ as some arbitrary function, then the
solution to (\ref{AH}) is
\be
A(t,x,y,z)\,=\,\exp\,[\int_0^t\,dt_1\,(B_y(t_1,x,y,z)\,
{\partial}_z\,-\,B_z(t_1,x,y,z)\,{\partial}_y)]
\,a_1(x,y,z),
\label{soloone}
\ee
where $a_1(x,y,z)$ is the value of $A$ at $t=0$ as in
(\ref{aitch}). The exponential here is defined by its power
series with the $n$'th term in this series being %
\ba
& &\int_0^t\,dt_1\,
\int_0^{t_1}\,dt_2 \dots
\int_0^{t_{n-1}}\,dt_n\, \nonumber\\
& &\left[
(B_y(t_1)\,{\partial}_z\,-\,B_z(t_1)\,{\partial}_y)\,
\dots
(B_y(t_n)\,{\partial}_z\,-\,B_z(t_n)\,{\partial}_y)
\right]\,
a_1(x,y,z).
\ea
We now must impose (\ref{BEE}) as a consistency condition on this
solution. This gives us
\ba
& & \hskip-1cm A(t,x,y,z)\,=\nonumber\\
& &
\hskip-1cm \exp\,\left[\int_0^t \,dt_1
\int_0^{t_1} \, dt_2\,
(A_{xy}(t_2,x,y,z)\,{\partial}_z -
A_{xz}(t_2,x,y,z)\,{\partial}_y) \right]
a_1(x,y,z).
\label{form}
\ea
Formally, this equation can now be solved iteratively. We can
make successive approximations
\be
A^{(0)}\,=\,a_1(x,y,z),
\ee
\be
A^{(1)}\,=\,\exp\,[
((t\,a_{0xy}\,+\,{t^2 \over{2}}a_{1xy})\,{\partial}_z\,
-\,
(t\,a_{0xz}\,+\,{t^2 \over{2}}a_{1xz})\,{\partial}_y)
\,]\,a_1(x,y,z),
\ee
\be
A^{(n+1)}\,=\,\exp\,[\int_0^t \,dt_1 \,
\int_0^{t_1} \, dt_2\,
(A^{(n)}_{xy}\,{\partial}_z\,-\,A^{(n)}_{xz}\,{\partial}_z\,)
\,]\,a_1(x,y,z),
\label{solve}
\ee
for $n\,\ge1$. Then defining $A\,=\,\lim_{n \to \infty}\,A^{(n)}$
gives the formal solution for $A$%
\footnote{It is beyond the scope of this paper to show that the
$A^{(n)}$ actually do converge to a well defined limit.}. %
Integrating $A$ with respect to $t$ and imposing $h(t=0) =
a_0(x,y,z)$ then gives us a solution of (\ref{biggy}).

Finally, we note that (\ref{AH}) means that the quantity
$A(t,x,{\tilde y},{\tilde z})$ is $t$ independent, where
${\tilde y}$ and ${\tilde z}$ are defined implicitly by
\be
{\tilde y}(t)\,=\,y\,+\,\int_0^{t}\,dt_1\,
B_z(t_1,x,{\tilde y}(t_1),{\tilde z}(t_1)),
\ee
\be
{\tilde z}(t)\,=\,z\,-\,\int_0^{t}\,dt_1\,
B_y(t_1,x,{\tilde y}(t_1),{\tilde z}(t_1)).
\ee
This may be important if we were to look for action angle
variables for the system. It also implies that $A(t,x,y,z) =
a_1(x,y',z')$, where the coordinates $y',z'$ are defined by
${\tilde y}(t,x,y',z') = y, {\tilde z}(t,x,y',z') = z$. Thus the
dynamics are characterised by a coordinate transformation in
the $y,z$ plane%
\footnote{An earlier version of this paper
incorrectly stated that this transformation was area
preserving.}.

\section{Group Methods}

Several powerful techniques have been developed for the study of
{\pdes} \cite{ZWILL}. One of the most powerful is that of group
analysis \cite{OLVER,BLKU}. By studying the Lie algebra under
which a given system of {\pdes} is invariant, we can hopefully
find new solutions to these equations. One method of doing so is
to look for similarity solutions which are left invariant by the
action of some sub-algebra of this symmetry algebra. This will
reduce the number of independent variables present in the
equation, possibly reducing a partial differential equation to
an ordinary differential equation. However, such similarity
solutions, by construction, will have some symmetries imposed
upon them, so this method is not very useful if one is looking
for the general solution to a system of equations.

A more powerful method is to exponentiate the infinitesimal
action of the Lie algebra into a group action, which takes one
solution of the equation to another. However, even if this is
possible, it is unlikely that the group action can be used to
find the general solution to the equation from any given
solution.

Instead of attempting to find the symmetry algebra of
(\ref{biggy}), it is easier to work with the equivalent system
(\ref{AH}) and (\ref{BEE}). We find that (\ref{AH}) and
(\ref{BEE}) admit a symmetry group defined by the infinitesimal
generators
\be
{\bf \xi}_1\,=\,f_A\,{\partial}_t\,-\,f_x\,
{\partial}_B,
\label{gone}
\ee
\be
{\bf \xi}_2\,=\,(t\,g_{x}\,+\,B\,g_{A})_{A}\,
{\partial}_t\,+\,g_A\,{\partial}_x\,-\,g_x{\partial}_A\,-\,
(t\,g_{x}\,+\,B\,g_{A})_{x}\,{\partial}_B,
\label{gtwo}
\ee
\be
{\bf \xi}_3\,=\,k\,t\,{\partial}_t\,+\,
k\,x\,{\partial}_x\,+\,k\,y\,{\partial}_y,
\label{gthree}
\ee
\be
{\bf \xi}_4\,=\,l_z\,{\partial}_y\,-\,
l_y\,{\partial}_z,
\label{gfour}
\ee
where $f$ and $g$ arbitrary functions of $x$ and $A$, $l$ is an
arbitrary function of $y$ and $z$, and $k$ is an arbitrary
constant. ${\bf \xi}_3$ just generates dilations, whereas ${\bf
\xi}_4$ generates area preserving diffeomorphisms in the
$y-z$ plane. Although ${\bf \xi}_4$ gives a representation of
$W_{\infty}$ (modulo cocycle terms) \cite{BAKAS}, they are
really only coordinate transformations, so are not too
interesting. However, we have two interesting symmetries,
generated by ${\bf \xi}_1$ and ${\bf \xi}_2$.

It is possible to exponentiate the action of ${\xi}_1$ directly
for an arbitrary function $f$. We find that if $A\,(t,x,y,z)$ and
$B\,(t,x,y,z)$ are a solution of the system (\ref{AH}) and
(\ref{BEE}) then we can implicitly define a new solution,
${\tilde A}$ and ${\tilde B}$, by %
\be
{\tilde A}\, =\,
A\,(t\,+\,f_A\,(x,{\tilde A}),\,x,\,y,\,z),\,
{\tilde B}\,=\,
B(t\,+\,f_A\,(x,{\tilde A}),\,x,\,y,\,z)\,+\,f_x(x,{\tilde A}),
\ee
for any function $f\,(x,A)$. Using this implicit form we can
solve iteratively for the functions ${\tilde A}$ and ${\tilde
B}$ given functions $A,B$ and $f$. This means that given one
solution of (\ref{biggy}), we can form an infinite dimensional
family of solutions depending on that solution. For a given
function $g$ we can also exponentiate the action of ${\xi}_2$,
although its action cannot be exponentiated directly for a
general function $g$.

Although both (\ref{gthree}) and (\ref{gfour}) give rise to
infinite dimensional families of solutions from any given
solution, they are not enough to derive a solution with
arbitrary initial data from any given solution.

If we compute the commutators of generators ${\bf \xi}_1(f_i)$
and ${\bf \xi}_2(g_j)$ for arbitrary functions $f_i$ and $g_j$
we find that they obey the algebra
\be
[{\bf \xi}_1 (f_1), {\bf \xi}_1 (f_2)]\,=\,0,
\ee
\be
[{\bf \xi}_1 (f), {\bf \xi}_2 (g)]\,=\,
{\bf \xi}_1(f_A\,g_x\,-\,f_x\,g_A),
\ee
\be
[{\bf \xi}_2 (g_1), {\bf \xi}_2 (g_2)]\,=\,
{\bf \xi}_2(g_{1A}\,g_{2x}\,-\,g_{1x}\,g_{2A}).
\ee

If we define a basis for transformations
\be
{}^{(\alpha)}T^{m}_{i} = {\bf \xi}_{\alpha}(x^{i+1}\,A^{m+1})
\ee
for $\alpha = 1,2$, where $m$ and $i$ are integers, then the
above algebra becomes
\be
[{}^{(1)}T^{m}_{i},{}^{(1)}T^{n}_{j}]\,=\,0,
\label{commute}
\ee
\be
[{}^{(1)}T^{m}_{i},{}^{(2)}T^{n}_{j}]\,=\,
((m+1)(j+1)\,-\,(n+1)(i+1)){}^{(1)}T^{m+n}_{i+j},
\ee
\be
[{}^{(2)}T^{m}_{i},{}^{(2)}T^{n}_{j}]\,=\,
((m+1)(j+1)\,-\,(n+1)(i+1)){}^{(2)}T^{m+n}_{i+j}.
\label{symp}
\ee
The algebra (\ref{symp}) is the algebra of locally area
preserving diffeomorphisms which, modulo cocycle terms, is
just the extended conformal algebra $W_{\infty}$ \cite{BAKAS}.
Thus (\ref{commute}) - (\ref{symp}) represent some
generalisation of $W_{\infty}$. These are similar results to
those found in \cite{PARK}.

We now note that equation (\ref{biggy}) can be
derived from the Lagrangian
\be
S\,=\,\int\,d^4 x \{ {1 \over 2}\,h_t^2\,+\,{1
\over 3}\, h_t\,(h_y\,h_{xz}\,-\,h_z\,h_{xy}) \}.
\ee
The Hamiltonian is then
\be
H\,=\,{1 \over 2}\int_{\Sigma}\,d^3x\,
(\pi - {1 \over 3} (h_y h_{xz} - h_z h_{xy}))^2.
\ee
where $\pi=h_t+{1\over3}(h_y h_{xz}-h_z h_{xy})$ is the
momentum canonically conjugate to $h$. We now define the
Poisson Bracket of functionals of $h$ and $\pi$ by
\be
\{ \alpha,\beta \} = \int_{\Sigma}\,d^3x\,
\left(
{{\delta \alpha}\over{\delta h}}{{\delta
\beta}\over{\delta \pi}}
-
{{\delta \alpha}\over{\delta \pi}}{{\delta
\beta}\over{\delta h}}\right),
\ee

The algebra (\ref{commute}) - (\ref{symp}) now reflects the fact
that we have two infinite families of conserved quantities%
\footnote{We are assuming that we can ignore surface terms.} %
of the form
\be
I(f(x,A))\,=\,\int_{\Sigma}\,f(x,A)\,d^{3}x
\ee
\be
I_2(g(x,A))\,=\,\int_{\Sigma}\,(t\,g_x\,+\,B\,g_A)\,d^3x.
\ee
These quantities all have vanishing Poisson brackets, i.e. they
are in involution. (The fact that these quantities are time
independent comes from the conservation equations
\be
{\partial}_t\,(f)\,+\,{\partial}_y\,(f\,B_z)\,-\,
{\partial}_z\,(f\,B_y)\,=\,0,
\ee
and
\ba
& &{\partial}_t\,(t\,g_x\,+\,B\,g_A)\,+\,
{\partial}_x\,(g)\,-\,
{\partial}_y\,(t\,g_{xA}\,B_z \,+\,g_A\,B\,B_z) \nonumber\\
& & \hskip2cm +\,
{\partial}_z\,(t\,g_{xA}\,B_y \,+\,g_A\,B\,B_y)\,=\,0,
\ea
which follow from (\ref{AH}) and (\ref{BEE}).)

\section{Solutions}

We begin by looking for solutions that admit a {\it
triholomorphic} Killing vector, ${\bf \xi}$. This means the
three complex structures, ${\bf J}^{i}$, are invariant under the
action of ${\bf \xi}$, i.e. ${\cal L}_{\xi}\,{\bf J}^{i}\,=\,0$,
where ${\cal L}$ is the Lie derivative. Using the relationship
between the complex structures and the vectors ${\bf V}_i$ given
in Section $2$ and the fact that $\xi$ is a Killing vector, we
see that we require ${\cal L}_{\xi}{\bf V}_i\,=\,0$.

If ${\partial}_x$ is a triholomorphic Killing vector, this means
that ${\partial}_x{\bf X}\,=\,{\partial}_x{\bf W}\,=\,0$, where
${\bf X}$ and ${\bf W}$ are as in (\ref{wexp}) and (\ref{xexp}).
This means $h$ is of the form $a(t,y,z)\,+\,x\,b(y,z)$ for some
functions $a$ and $b$. In terms of functions $A\,=\,h_t$ and
$B\,=\,h_x$ this means that $A\,=\,A(t,y,z),\,B\,=\,B(y,z)$, so
(\ref{BEE}) is automatically satisfied. If we take
$A(t=0)\,=\,a_1\,(y,z)$ and $B(t=0)\,=\,\phi\,(y,z)$, it is
straightforward to show that the solution to (\ref{AH}) is then
\be
A\,(t,y,z)\,=\,
\exp \{ t\,
({\phi}_y\,{\partial}_z\,-\,{\phi}_z\,{\partial}_y)\}
\,a_1\,(y,z),\,
B\,(y,z)\,=\,\phi\,(y,z).
\ee
For given functions $\phi$ and $a_1$ it is straightforward to
do the exponentiation, giving $A$ explicitly. Using the
exponentiated form of (\ref{gone}) and (\ref{gtwo}), we could now
use these solutions to generate new solutions which had
some restricted $x$-dependent initial data as well.

We can also consider metrics with a triholomorphic Killing
vector, ${\partial}_z$. This means we require ${\partial}_z{\bf
V}_i\,=\,0$. In this case, we take
$h\,=\,-\,t\,z\,+\,g\,(t,x,y)$. We then recover the result
\cite{GWGSWH} that $g$ must satisfy the three dimensional Laplace
equation $g_{tt}\,+\,g_{xy}\,=\,0$. The general solution to this
is known, and can be written in terms of two arbitrary functions
$a_0(x,y)$ and $a_1(x,y)$. An almost identical reduction occurs
if we take ${\partial}_y$ as a triholomorphic Killing vector.
Again, using the symmetries (\ref{gone}) and (\ref{gtwo}), we
can generate infinite dimensional families of new solutions,
that in general have no Killing vectors.

We note in passing that the solution corresponding to the
multi-centre Eguchi-Hansen metric \cite{GWGSWH} is
\be
A\,=\,-\,z\,+\,\alpha\,\sum_{i=1}^{s}\,
\sinh^{-1}\,\left({(t-t_i)\over{2
\sqrt{(x-x_i)(y-y_i)}}}\right),
\ee
\be
B\,=\,-\,{\alpha \over 2} \,\sum_{i=1}^{s}
{\sqrt{(t-t_i)^2\,+\,4\,(x-x_i)\,(y-y_i)}\over{(x-x_i)}}.
\ee
where $\alpha$ is a constant. This is the only metric with a
triholomorphic Killing vector that has a non-singular real
(euclidean) section \cite{HIT}.

At present, work is underway to complete a study of holomorphic
Killing vectors and to see how the problem relates to the known
results on such metrics \cite{BANDF,GANDD}.

\section{Conclusion}

We have shown how, at least formally, to construct the general
complex metric with {\sd} Riemann tensor. We have also studied
the symmetry algebra of the system and found two infinite
dimensional families of conserved quantities that have
vanishing Poisson brackets.

It should be emphasised that all the considerations here have
been inherently local in nature, and we have imposed no sorts of
boundary conditions on our solutions at infinity. If we were to
look for metrics that are well defined globally, this would lead
us to cohomological problems \cite{WANDW}, which appear to be
best tackled using the twistor formalism \cite{NLG}.

\section*{\bf Acknowledgements}

I would like to thank my supervisor, Stephen Hawking, for much
help and guidance in the course of this work. Also I thank
Andrew Dancer and Gary Gibbons for many helpful discussions on
{\hk} geometry.

This work was carried out under a Science and Engineering
Research Council (SERC) studentship.

\appendix
\section{The First Heavenly Equation}

Starting with (\ref{AH}) and (\ref{BEE}), instead of looking on
$A$ as a function of $t,\,x,\,y$ and $z$ we take $A$ as a
coordinate and look on $f \equiv t$ and $g \equiv B$ as
functions of $p \equiv A,\, q \equiv x,\, r \equiv y,\, s \equiv
z$. This transformation is well defined as long as $A_t \neq 0$.
Inverting (\ref{AH}) and (\ref{BEE}) gives
\be
f_q\,=\,-\,g_p,
\label{ahtwo}
\ee
\be
f_r\,g_s\,-\,f_s\,g_r\,=\,1.
\label{beetwo}
\ee
(\ref{ahtwo}) means we can introduce a function $\Omega
(p,q,r,s)$ such that $f\,=\,-\,{\Omega}_p,\,g\,=\,{\Omega}_q$.
(\ref{beetwo}) then means that $\Omega$ must satisfy %
${\Omega}_{ps}\,{\Omega}_{qr}\,-\,{\Omega}_{pr}\,{\Omega}_{qs}\,
=\,1$. %
Carrying out the same transformation on the line element
(\ref{metric}), we find it becomes %
$ds^2\,=\,{\Omega}_{pr}\,dp\,dr\,+\,{\Omega}_{ps}\,dp\,ds\,
+\,{\Omega}_{qr}\,dq\,dr\,+\,{\Omega}_{qs}\,dq\,ds$. %
Thus we have recovered the {\P} formalism.

\end{document}